# Long-Range Reading of Multiple Chipless Sensors from the Isoline Processing of 3D Radar Images

A. HADJ DJILANI[(1,3)], D. HENRY[(1)], A. EL SAYED AHMAD[(2)], P. PONS[(1)] and H. AUBERT[(1,3)]

(1) LAAS-CNRS, 31400 Toulouse, France, {ahadjdjila, dhenry, aelsayedah,ppons ,aubert}@laas.fr.
(2) Capgemini Engineering, Toulouse, France, ahmad.el-sayed-ahmad@capgemini.com.
(3) Toulouse University, Toulouse, France.

*Abstract*— In this paper, we report the long-range and wireless interrogation of multiple chipless sensors from the isoline processing of three-dimensional polarimetric radar images. A Frequency-Modulated Continuous-Wave Radar operating at 24 GHz is used for the indoor interrogation of four sensors in the basement of a Laboratory. In such cluttered environment, the proposed radar image processing based on isolines computation allows the wireless measurement range of sensors up to 5.8m.

## I. Introduction

Chipless sensors (i.e., sensors without batteries and integrated circuits) may offer advantages in comparison with active sensors for applications where electronic components may be damaged (e.g., in harsh environments) or where batteries cannot be replaced easily (e.g., in hard to reach areas). Chipless RFID (Radio Frequency Identification) sensors and SAW (Surface Acoustic Wave) sensors have been extensively studied, but they are often limited by their short reader-to-sensor distances (typically 1 or 2 m) [1]. To enlarge the reader-to-sensor distance, the signal-to-noise ratio can be increased by increasing the RF input power of the reader, but RFID applications are limited by regulations on the EIRP (Effective Isotropic Radiated Power). Another well-known solution consists of using depolarizing sensors to reduce the level of the electromagnetic clutter [2]. The Authors of the present study have recently reported the remote reading of a chipless sensor by using a millimeter-wave FM-CW (Frequency-Modulated Continuous-Wave) radar. Long reader-to-sensor distances (up to 58m) have been reached in a corridor from the radar interrogation of a depolarizing and retro-directive chipless sensor [3].

In this paper, we propose to extend the radar-based approach to the long-range wireless interrogation of multiple chipless sensors by using a 24 GHz Frequency-Modulated Continuous-Wave radar. Four sensors are located indoor at different radar-to-sensor separation distances. The sensors are similar to those used in [4] except that they are completely manufactured using three-dimensional 3D metallic additive technology. The original processing of 3D radar images based on isolines is applied here to remotely determine the dynamic range (or equivalently the ratio between the highest and smallest echo levels) of these sensors up to 5.8m.

## II. Radar-based Reading System

Passive sensors are interrogated using FM-CW radar (model DK-sR-1030e from IMST GmbH [5]) operating at 23.8GHz (wavelength of 12.6mm) with the bandwidth of 2GHz. The transmitting (Tx) channel of the radar is connected to a vertically-polarized (V-polarized) lens horn antenna (with high gain of 28dBi and half-power beamwidth of 6°) through a 60cm coaxial cable (see Figure 1). The two receiving channels Rx,1 and Rx,2 are respectively connected to V-polarized and H-polarized rectangular horn antennas (gain of 20dBi). In this study, VV refers to the co-polarized configuration, that is, the configuration for which the signal backscattered from the scene is received by the V-polarized Rx2 antenna. In addition, VH refers to the cross-polarized configuration for which the backscattered signal is received by the H-polarized Rx1 antenna.

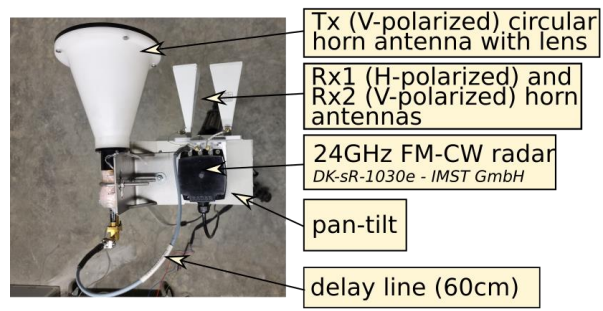

Figure 1. Top view of the radar reading system

The radar reading system with antennas is mounted on a pan-tilt to perform the mechanical scanning of the radar beam with an angular step of around 0.3° in azimuth (angle $\phi$) and 1° in elevation (angle $\theta$). In each direction ($\theta$, $\phi$) the radar delivers a beat frequency spectrum for the VV and VH configurations, that is, the variation of radar echo level as a function of the distance in co- and cross-polarizations. From the set of beat frequency spectra obtained in all directions, we can build 3D polarimetric radar images in elevation / azimuth / range coordinates ($\theta$, $\phi$, R).

## III. Remote Interrogation of Multiple Chipless Sensors

Four depolarizing and chipless sensors, denoted by sensors #1, #2, #3 and #4, are located in the basement of our Laboratory for reader-to-sensor distances of respectively 2.5m, 3.6m, 4.4m and 5.8m (see Figure 2). In this indoor environment, metal grids attached to the ceiling, the concrete pillars, the floor, the platform supporting the sensors and cables may contribute to the electromagnetic clutter. The sensors are similar to those used in [4] except that they are completely manufactured using 3D metallic additive technology. In the experiment, they are either in ON or OFF state. In ON state, the signal backscattered by the sensor is primarily H-polarized and consequently, radar echo level of the sensor should be significantly lower in VV configuration than in VH configuration. In OFF state, the backscattered signal is mainly vertically-polarized and therefore, the radar echo level of the sensor is expected to be much higher in VV configuration then in VH configuration. As shown in the next paragraph, the ratio between the smallest and highest echo levels at a given reader-to-sensor separation distance allows defining the dynamic range of the sensor at this distance.

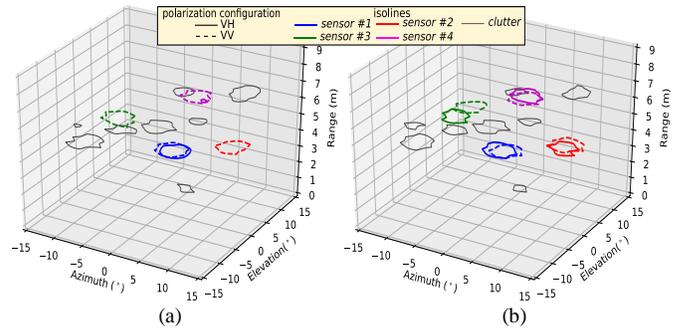

(a) (b)

Figure 3. Computed isolines in ($\theta$, $\phi$, R) coordinates system in (a) scenario 1, and (b) scenario 2. Solid and dashed lines represent isolines computed respectively in VH and VV configurations.

TABLE I. DYNAMIC RANGE OF THE SENSORS IN VV AND VH CONFIGURATIONS AT THEIR RESPECTIVE DISTANCES FROM THE RADAR

| sensor | R (m) | $e_{max}^{VV}$ (dB) | | | $e_{max}^{VH}$ (dB) | | |
|---|---|---|---|---|---|---|---|
| | | ON | OFF | $\Delta e_{max}^{VV}$ | ON | OFF | $\Delta e_{max}^{VH}$ |
| #1 | 2.5 | -1.9 | 3.7 | 5.6 | 2.6 | -4.9 | 7.5 |
| #2 | 3.6 | -4.9 | -3.0 | 1.9 | -0.9 | -17.3 | 16.4 |
| #3 | 4.5 | -0.1 | -1.0 | 0.9 | 1.1 | -16.2 | 17.3 |
| #4 | 5.8 | -4.5 | -1.6 | 2.9 | 0.2 | -9.7 | 9.9 |

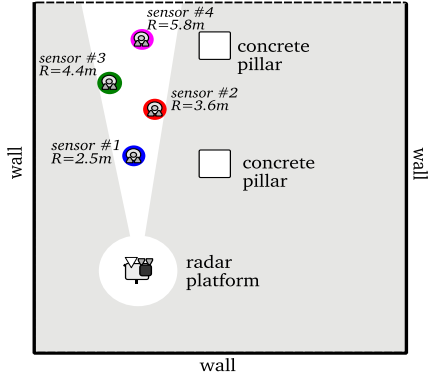

Figure 2. Top view of the indoor environment used for remote interrogating four chipless sensors located at reader-to-sensor distances of 2.5m, 3.6m, 4.4m and 5.8m.

We propose here two scenarios with various combinations of ON and OFF states of four chipless sensors: in the first scenario, all sensors are in OFF state, in the second scenario, all sensors are in ON state. From the 3D polarimetric radar images obtained from each scenario, we derive the isolines, that is, lines in the 3D images along which the radar echo level is the same. Isolines are computed from the algorithm detailed in [6] and are displayed in Fig. 3 in ($\theta$, $\phi$, R) coordinates system for the two scenarios in VV and VH configurations and for a minimal threshold of echo level of -10dB. Isolines of the sensors in OFF states are not all computed in VH configuration, as the radar echo level is smaller than -10dB. This is specifically the case for sensors #2 and #3 in the first scenario. Inversely, VH isolines in ON state are displayed for both scenarios because the radar echoes of sensors are sufficiently high (>-10dB) in this case. Moreover, isolines in the VV configuration are computed in the VH configuration both in ON and OFF states at same positions of isolines. Their shape differs because VV isolines originate from the structural backscattering mode of the sensors, whereas VH isolines are generated by the sensing backscattering mode.

The dynamic range $\Delta e_{max}^{VV}$ (resp. $\Delta e_{max}^{VH}$) of a sensor in VV (resp. VH) configuration is defined as the ratio between the highest echo level $e_{max}^{VV}$ (resp. $e_{max}^{VH}$) inside isolines in VV (resp. VH) configuration when the sensor is ON and the highest echo level in this configuration when the sensor is OFF state. Highest radar echo levels for both polarization configurations as well as the resulting dynamic range of the four sensors at their respective distance from the radar are reported in Table I. As expected, we observe that the dynamic range is significantly larger in VH configuration than in VV configuration. For instance, $\Delta e_{max}^{VV}$ = 0.9dB in VV polarization while $\Delta e_{max}^{VH}$ = 17.3dB in VH configuration for sensor #3. This result confirms that the cross-polarized configuration allows increasing the signal-to-noise ratio and mitigating the electromagnetic clutter.

## IV. CONCLUSION

We have demonstrated in this study the feasibility of interrogating multiple (four) chipless sensors at distances up to 5.8m from the processing of polarimetric radar images. Future work will involve characterizing the sensitivity of the radar measurement for larger reader-to-sensor separation distances (tenth of meters), as well as identifying each sensor from their respective Radar echoes.

## V. ACKNOWLEDGEMENT

The authors thank the National Agency for Research (France) for financial support through the S²LAM Project.